\documentclass[aps,prb,twocolumn,groupedaddress]{revtex4}
\usepackage{graphicx}

\begin{document}

\noindent \textbf{\large{Direct electronic measurement of the spin
Hall effect}}

\vspace{5mm}

\noindent \textsf{\textbf{S.O. Valenzuela* and M. Tinkham}}

\vspace{5mm}

\noindent \textsl{\small{Department of Physics, Harvard
University, Cambridge, MA 02138, USA.}}

\vspace{5mm}

\textbf{The generation, manipulation and detection of
spin-polarized electrons in nanostructures define the main
challenges of spin-based electronics\cite{wolf}. Amongst the
different approaches for spin generation and manipulation,
spin-orbit coupling, which couples the spin of an electron to its
momentum, is attracting considerable interest. In a
spin-orbit-coupled system a nonzero spin-current is predicted in a
direction perpendicular to the applied electric field, giving rise
to a ``spin Hall effect"\cite{dya1,dya2,hirsch}. Consistent with
this effect, electrically-induced spin polarization was recently
detected by optical techniques at the edges of a semiconductor
channel\cite{aws1} and in two-dimensional electron gases in
semiconductor heterostructures\cite{wund,aws2}. Here we report
electrical measurements of the spin-Hall effect in a diffusive
metallic conductor, using a ferromagnetic electrode in combination
with a tunnel barrier to inject a spin-polarized current. In our
devices, we observe an induced voltage that results exclusively
from the conversion of the injected spin current into charge
imbalance through the spin Hall effect. Such a voltage is
proportional to the component of the injected spins that is
perpendicular to the plane defined by the spin current direction
and the voltage probes. These experiments reveal opportunities for
efficient spin detection without the need for magnetic materials,
which could lead to useful spintronics devices that integrate
information processing and data storage.}

The spin Hall effect (SHE), which was first described by Dyakonov
and Perel\cite{dya1,dya2} and more recently by
Hirsch\cite{hirsch}, was proposed to occur in paramagnetic
materials as a consequence of the spin-orbit interaction. In
analogy to the standard Hall effect, the SHE refers to the
generation of a pure spin current transverse to an applied
electric field that results in an accompanying spin imbalance in
the system. These early theoretical studies considered an
extrinsic\cite{dya1,dya2,hirsch,zhang1} SHE originating from an
asymmetric scattering for spin-up and spin-down electrons. It was
pointed out that, after scattering off an impurity, there is a
spin-dependent probability difference in the electron trajectories
which generates the spin imbalance. In the recently introduced
intrinsic\cite{zhangsc,sin1} SHE, spin imbalance is expected to
occur even in the absence of scattering as a result of the band
structure.

Several experimental schemes have been proposed to electrically
detect the extrinsic SHE in metals\cite{hirsch,zhang1,bra1,bra2}.
However, these schemes are difficult to implement. Spin-related
phenomena such as anisotropic magnetoresistance (AMR) in
ferromagnetic (FM) electrodes, spin dependent interface
scattering, or standard and anomalous Hall effects could render
the SHE signal unobservable. We use the measurement scheme in Fig.
1 to study the SHE isolated from such spurious phenomena.

\begin{figure}[b]
\vspace{-5mm}
\includegraphics[width=3.0in]{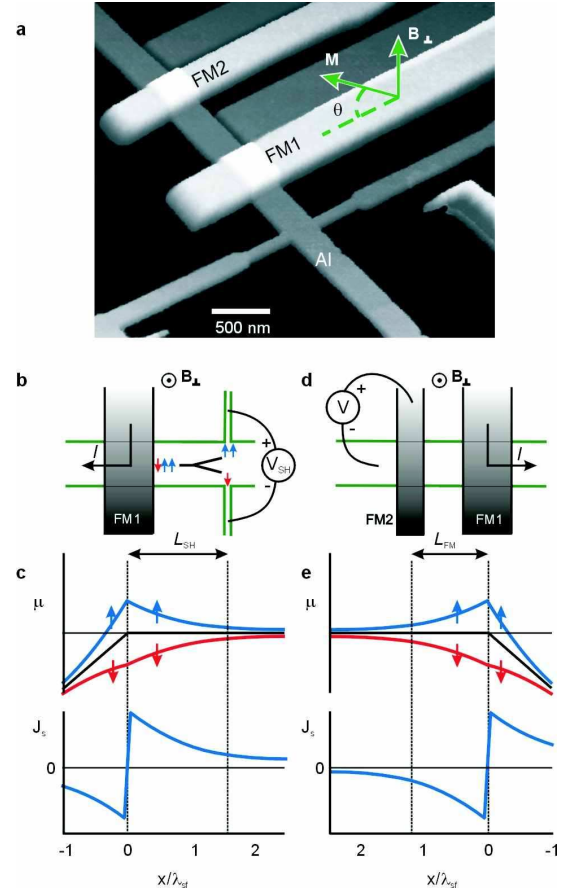}
\vspace{-5mm} \caption{Geometry of the devices and measurement
schemes. (a) Atomic force microscope image of a device. A thin
aluminum (Al) Hall cross is oxidized and contacted with two
ferromagnetic electrodes with different widths (FM1 and FM2). (b)
Spin Hall measurement. A current $I$ is injected out of FM1 into
the Al film and away from the Hall cross. A spin Hall voltage,
$V_{SH}$, is measured between the two Hall probes. $V_{SH}$ is
caused by the separation of up and down spins due to spin-orbit
interaction in combination with a pure spin current. (c) Top:
spatial dependence of the spin-up and spin-down electrochemical
potentials, $\mu_{\uparrow, ~\downarrow}$. The black line
represents the electrochemical potential of the electrons in the
absence of spin injection. Bottom: associated spin current, $J_s$.
The polarized spins are injected near $x=0$ and diffuse in both Al
branches in opposite directions.  The sign change in $J_s$
reflects the flow direction. (d) Spin-transistor measurement for
device characterization. $I$ is injected out of FM1 into the Al
film and away from FM2. A voltage $V$ is measured between FM2 and
the left side of the Al film. (e) Same as in (c).} \label{fig1}
\end{figure}

It is natural to expect that, if a \emph{charge}-current induces a
transverse \emph{spin}-imbalance through the spin-orbit
interaction, a \emph{spin}-current will induce a transverse
\emph{charge}-imbalance (and a measurable voltage) by the same
mechanism\cite{hirsch}. In our experiments (Fig. 1a), a
ferromagnetic electrode (FM1) is used to inject spin-polarized
electrons via a tunnel barrier in one of the arms of an aluminum
(Al) Hall cross. As shown in Fig. 1b, the injected current $I$ is
driven away from the Hall cross, where only a pure spin-current
flows as a result of the spin injection. If spin-orbit scattering
is present, the spin-current could induce a transverse spin-Hall
charge imbalance and generate a measurable voltage, $V_{SH}$. At a
distance $L_{SH}$ from the spin injector, the two transverse
probes forming the Hall cross are used to measure such a voltage.
The tunnel barrier\cite{vanwees1} is important because it assures
a uniformly distributed injection current, and enhances the
polarization of the injected electrons\cite{SOV1,SOV2}. The
non-local nature of our measurement scheme eliminates spurious
effects due to AMR in the FM electrode or the anomalous and
standard Hall effects. As no net charge-current flows into the
Hall cross, the direct generation of voltage by the standard Hall
effect is precluded. Located at a distance $L_{FM}$ from FM1 is a
second ferromagnetic electrode (labelled FM2 in Fig. 1a) that is
used for sample characterization, as explained below.

We prepare the devices with electron beam lithography and a
two-angle shadow-mask evaporation technique to produce tunnel
barriers \emph{in situ}. The Al cross, with arms 400 nm and 60 nm
wide, is first deposited at normal incidence onto a Si/SiO$_2$
substrate using electron beam evaporation. Next, the Al is
oxidized in pure oxygen (150 mTorr for 40 min) to generate
insulating Al$_2$O$_3$ barriers. After the vacuum is recovered,
the two FM electrodes (400 and 250 nm wide, and 50 nm thick) are
deposited under an angle of 50$^{\circ}$, measured from the normal
to the substrate surface. The FM electrodes form tunnel junctions
where they overlap with the Al strip with a typical tunnel
resistance of 4 k$\Omega$ for FM1 and 6.5 k$\Omega$ for FM2. For
the FM electrodes, we use CoFe (80 wt. \% Co), which provides a
large polarization when combined with Al$_2$O$_3$ as a tunneling
barrier\cite{SOV1,douwe}. The difference in the FM electrodes
widths is required to get different coercive fields. A crucial
point in our fabrication procedure is that no image of the Hall
cross is deposited during the evaporation of FM1 and FM2, as
observed in the atomic force micrograph of Fig. 1a. The FM
deposits on the wall of the lithography mask and it is removed by
lift-off, preventing unwanted short circuits (see Ref.
\onlinecite{SOV1} and Supplementary Information, Fig. 1).

The spin polarization $P$ of the electrons injected by FM1 depends
on the effective tunnel conductances for spin-up and spin-down
electrons, $G_{\uparrow}$ and $G_{\downarrow}$, and can be written
as
$P=(G_{\uparrow}-G_{\downarrow})/(G_{\uparrow}+G_{\downarrow})$.
The spin-polarized current causes unequal electrochemical
potentials for the spin up and down populations in Al (Fig 1c, top
panel) and a spin current which flows to both sides of the contact
and decays with the spin diffusion length, $\lambda_{sf}$ (Fig.
1c, bottom panel). As a result, the charge imbalance that is built
up at the edges of the Al strip, and $V_{SH}$ are expected to be
proportional to $P$ and to decay with $\lambda_{sf}$. Their
magnitude is determined by the anomalous Hall
operator\cite{zhang1,bra1,bra2,zhang2}, $\sigma_{SH} ~
\mathbf{\hat{\sigma}} \times \mathbf{E}^{\sigma}$, where
$\sigma_{SH}$ denotes the spin Hall conductivity, $\sigma$ is the
spin index, and $\mathbf{E}^{\sigma}$ is an effective
spin-dependent ``electric" field, which follows from the
spin-dependent electrochemical potential $\mu^{\sigma}$ along the
Al strip, i.e. $\mathbf{E}^{\sigma}(\mathbf{r})=-\nabla
\mu^{\sigma}(\mathbf{r})$.

Spin imbalance in the Al film occurs with a defined spin direction
given by the magnetization orientation of the FM electrode.
Consequently, $V_{SH}$ is expected to vary when a magnetic field
perpendicular to the substrate, $B_{\perp}$, is applied and the
magnetization $\mathbf{M}$ of the electrode is tilted out of the
substrate plane. Defining $\theta$ as the angle between
$\mathbf{M}$ and the electrode axis (Fig. 1a), we see, from the
cross product in the anomalous Hall operator, that $V_{SH}$ is
proportional to $\sin \theta$, correlating with the component of
$\mathbf{M}$ normal to the substrate. As discussed below, $\theta$
can be set to be any value between $\sim -90$ and $\sim 90$
degrees with $|B_{\perp}| \lesssim $ 2T.

\begin{figure}[top]
\includegraphics[width=3.2 in]{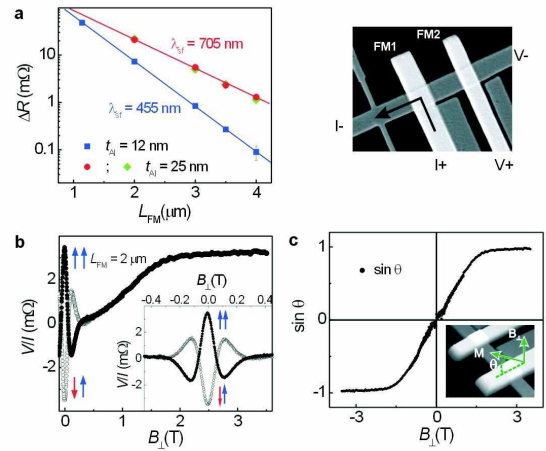}
\vspace{-5mm} \caption{Spin-transistor measurements and spin
precession. Top-right: scanning electron micrograph of the device
and the measurement scheme. (a) $\Delta R= \Delta V/I$ as a
function $L_{FM}$ for 3 sets of samples. Green diamonds and red
circles are for an Al thickness $t_{Al}$= 25 nm, blue squares are
for $t_{Al}$= 12 nm. The lines are best linear fits. For raw data,
see Supplementary Information, Fig. 2. (b) Transresistance change
due to spin precession as a function of $B_{\perp}$ (main panel,
up to 3.5 T and inset, up to 0.5 T). Results are symmetric about
$B_{\perp}=0$. The arrows indicate the relative orientation of the
magnetizations of FM1 and FM2. $t_{Al}$= 12 nm; $L_{FM}=2$ $\mu$m.
(c) $\sin \theta$ as a function of $B_{\perp}$ extracted from data
in (b). Inset: magnetization direction of the FM electrodes
relative to the substrate.} \label{fig2}
\end{figure}

We first characterize the properties of our device. FM1 and FM2,
together with the Al strip, define a reference Johnson and Silsbee
spin-transistor\cite{js1,js2,js3} with Jedema \textit{et al.}
thin-film layout\cite{vanwees1}. We use this spin-transistor to
obtain  $P$, $\lambda_{sf}$, and $\theta$ at $B_{\perp}\neq 0$.
Measurements are performed at 4.2 K as represented in Figs. 1d and
1e (note the change in the direction of $I$ in Figs. 1d and 1b).
Both $P$ and $\lambda_{sf}$ are obtained by measuring the spin
transresistance $\Delta R=\Delta V/I$ as a function of $L_{FM}$,
where $\Delta V$ is the difference in the output voltage between
parallel and antiparallel magnetization configurations of the FM
electrodes at zero magnetic
field\cite{vanwees1,js1,js2,js3,alex,havi}. A lock-in amplifier is
used with $I$ equal to 50 $\mu$A. At larger currents, a marked
decrease in the spin polarization is observed, as previously
reported\cite{SOV2}.

Figure 2a shows results for three different batches of samples
with different Al thickness, $t_{Al}$. As noted
recently\cite{SOV1}, in Al films $\lambda_{sf}$ is strongly
dependent on $t_{Al}$. We use this property to study the SHE in Al
with different $\lambda_{sf}$. The top two sets of measurements
correspond to $t_{Al}= 25$ nm (red circles and green diamonds),
whereas the bottom one (blue squares) corresponds to $t_{Al}= 12$
nm. The data shown in red and green are almost identical,
demonstrating the high reproducibility of our sample fabrication.
By fitting the data to\cite{vanwees1,js1,js2,js3,alex} $\Delta R
=P^2 \frac{\lambda_{sf}}{\sigma_{c}A}\exp(-L_{FM}/\lambda_{sf})$,
where $\sigma _{c}$ is the Al conductivity [$\sigma _{c}(12~
\mathrm{nm})=1.05 ~ 10^7 (\Omega \mathrm{m})^{-1}$; $\sigma
_{c}(25 ~\mathrm{nm})=1.7 ~ 10^7 (\Omega \mathrm{m})^{-1}$] and
$A$ the cross sectional area of the Al strip, we obtain $P =
0.28$, $\lambda_{sf}(12~ \mathrm{nm})= 455 \pm 15$ nm, and
$\lambda_{sf}(25 ~\mathrm{nm}) = 705 \pm 30$ nm.

The tilting angle can be obtained from the spin precession results
shown in Fig. 2b, which are measured with $B_{\perp}$ between -3.5
to 3.5 T. For small $B_{\perp}$, the magnetizations of the FM
electrodes remain in plane due to shape anisotropy, and the
measurements show the Hanle effect associated with precessing
spins. As $B_{\perp}$ increases, the magnetizations tilt out of
plane. For large enough $B_{\perp}$, they orient completely along
the field and the measurements saturate to a positive constant
value. The output voltage normalized by its value, $V_0$, at
$B_{\perp} = 0$ has the form $V_{\pm}/V_0=\pm f(B_{\perp})
\cos^2(\theta) + \sin^2(\theta)$, for initially parallel (+) and
antiparallel (-) configurations of the magnetizations, with
$f(B_{\perp})$ a given function of
$B_{\perp}$\cite{vanwees1,js2,js3}. By noting that $(V_+ +
V_-)/V_0 = 2 \sin^2 \theta$, $f(B_{\perp})$ is eliminated, and the
calculation of $\sin \theta$ is straightforward. The result of
this calculation is presented in Fig. 2c. At $B_{\perp}=0$,
$\theta=0$ due to shape anisotropy. When $B_{\perp}$ is applied,
the magnetization follows the Stoner-Wohlfarth model\cite{ohan}
with a saturation field $B_{\perp}^{sat}$ of about 1.55 T for
which $\sin \theta$ approaches one, and the magnetization aligns
with the field.

Having determined $P$, $\lambda_{sf}$ and $\theta$, we study the
SHE using the measurement configuration shown in Fig. 1b. A change
in $B_{\perp}$ results in a change in $V_{SH}$. By considering the
spin diffusion using a semiclassical Boltzmann approximation, the
spin Hall resistance $R_{SH}=V_{SH}/I$ at $L_{SH}$ (see Fig.1b)
can be calculated\cite{zhang1,zhang3}:

\begin{equation}
R_{SH}=\frac{\Delta R_{SH}}{2} \sin \theta,
\label{eq1}\end{equation}

\noindent with

\begin{equation}
\Delta R_{SH}=\frac{P}{t_{Al}} \frac{\sigma_{SH}}{\sigma_{c} ^2}
\exp(-L_{SH}/\lambda_{sf}). \label{eq2}\end{equation}

Here it is assumed that the distance between the two voltage
probes (Al strip width) is smaller than $\lambda_{sf}$. $\Delta
R_{SH}$ in equation (2) was obtained by Zhang (see Refs.
\onlinecite{zhang1} and \onlinecite{zhang3}). The factors $\sin
\theta$ and $\frac{1}{2}$ in equation (1) account for the
orientation of the spins and the fact that they diffuse away from
FM1 into two (Al) branches instead of one, as in Ref.
\onlinecite{zhang3}. Note that, in the present configuration, the
measurements are not affected by spin-precession because the
component of the spins perpendicular to the substrate (or the
transverse component of the anomalous velocity) is not modified by
this effect.

Figure 3 shows $R_{SH}$ as a function of $B_{\perp}$ for samples
with $L_{SH}$= 860, 590, and 480 nm (circles) and $t_{Al}= 12$ nm.
The sign convention is shown in the inset, where a positive
$B_{\perp}$ is pointing out of the page. $B_{\perp}$ is swept
between -3.5 and 3.5 T. We observe a linear response around
$B_{\perp}=0$, followed by a saturation on the scale of
$B_{\perp}^{sat}$, both for positive and negative $B_{\perp}$.
Similar results were obtained in about 20 samples with different
$t_{Al}$.

\begin{figure}[top]
\vspace{-14mm}
\includegraphics[width=3.2in]{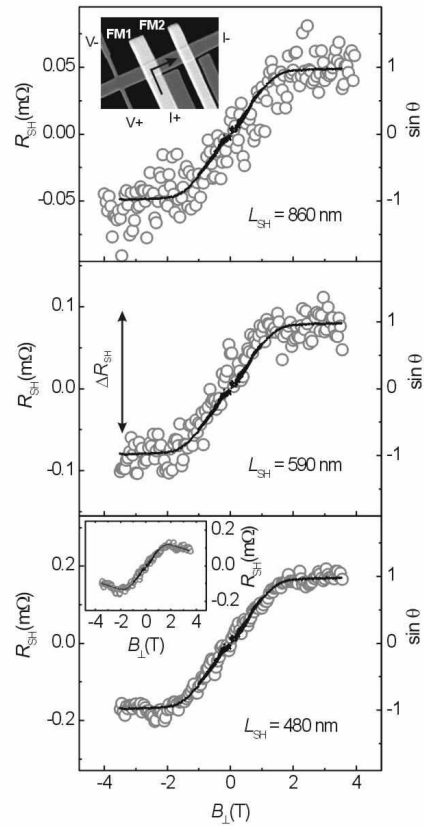}
\vspace{-16mm} \caption{Spin Hall effect. $R_{SH}$ vs.
$B_{\perp}$, for $t_{Al}= 12$ nm and $L_{SH}=860$ nm (top),
$L_{SH}=590$ nm (middle), and $L_{SH}=480$ nm (bottom). The top
panel inset shows a scanning electron micrograph of the device and
the measurement scheme with the sign convention. Top and middle
panels and inset in the bottom panel show raw data. In the bottom
panel, the linear background was subtracted from the data in the
inset. For comparison, the measured value of $\sin \theta$ is
shown in each panel. Note the decrease in $R_{SH}$ with $L_{SH}$.}
\label{fig3}
\end{figure}

The saturation in $R_{SH}$ for $|B_{\perp}|
> B_{\perp}^{sat}$ strongly suggests that the device output is related to the magnetization orientation
of the FM electrode (see Fig. 2c) and the spin-Hall effect. This
idea is further reinforced by comparing $R_{SH}(B_{\perp})$ with
the magnetization component perpendicular to the substrate, which
as discussed above is proportional to $\sin \theta (B_{\perp})$
(lines in Fig. 3). The agreement is excellent; the proportionality
between $R_{SH}$ and $\sin \theta$ in equation (1) is closely
followed.

For $L_{SH}$ smaller than $\lambda_{sf}$, a small linear term in
$R_{SH}(B_{\perp})$ becomes apparent (see bottom panel of Fig. 3
and inset). We believe that this term arises from the contribution
of the orbital Lorentz force to the spin-Hall induced charge
current\cite{bra2}, which is a second-order effect that decays
faster with $L_{SH}$ than the spin Hall voltage itself.

Measurements were again performed with $I$ equal to 50 $\mu$A.
Larger currents lead to a decrease in $R_{SH}$ as the voltage drop
at the tunnel junction increases. This correlates with the
decrease in the spin polarization observed with the
spin-transistor measurements\cite{SOV2} and further supports the
interpretation of our results as being spin-related.

\begin{figure}[t]
\vspace{-10mm}
\includegraphics[width=3.4in]{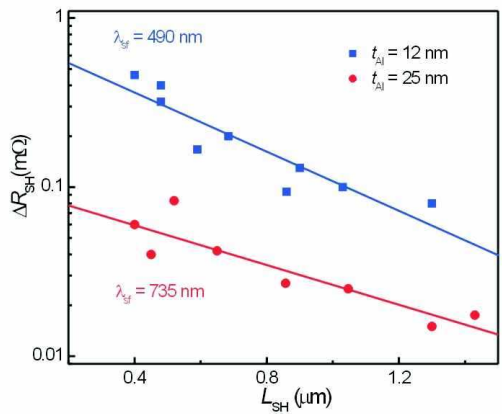}
\vspace{-30mm} \caption{Overall change of the spin-Hall
resistance, $\Delta R_{SH}$, between large negative and large
positive $B_{\perp}$. $\Delta R_{SH}$ is obtained by a best fit to
equation (1) using data in Fig. 3. Solid lines are best fits to
equation (2). Blue squares (red circles) are for $t_{Al} = 12(25)$
nm.} \label{fig4}
\end{figure}

Consistent with equation (2), the overall change of $R_{SH}$,
$\Delta R_{SH}$, decreases as a function of $L_{SH}$ (Fig. 3). By
fitting the magnetic field dependence of $R_{SH}$ in Fig. 3 to
equation (1), we have obtained $\Delta R_{SH}$ for each sample.
The results are shown in Fig. 4 as a function of $L_{SH}$, which
are then fitted to equation (2) in order to obtain $\lambda_{sf}$
and $\sigma_{SH}$. We find $\lambda_{sf}$(12 nm) = $490\pm 80$ nm,
and $\lambda_{sf}$(25 nm) = $735 \pm 130$ nm, in good agreement
with the values obtained independently with the control
spin-transistor (Fig. 2a). By using $P=0.28$, we find $\sigma
_{SH}$(12 nm) = $(3.4 \pm 0.6)~10^3$ ($\Omega$ m)$^{-1}$ and
$\sigma _{SH}$(25 nm) = $(2.7 \pm 0.6)~10^3$ ($\Omega$ m)$^{-1}$.
The ratio $\sigma _{SH}/\sigma_{c} \sim$ 1-3 $10^{-4}$ is of the
same order of magnitude as the one obtained
experimentally\cite{aws1} and theoretically\cite{bert,tse} for
GaAs. The predicted $\sigma _{SH}$, when considering $\delta$-like
scattering centers, is $\alpha\hbar e^2 N_0 /3m$
\cite{zhang1,bra1}, where $\hbar$ is Planck's constant divided by
$2\pi$, $N_0 = 2.4~ 10^{28}$ states/eV m$^3$ is the density of
states of Al at the Fermi energy \cite{pap}, $\alpha \sim 0.006$
is the dimensionless spin-orbit coupling constant of Al
\cite{bra2}, and $e$ and $m$ are the charge and mass of the
electron. Without free parameters, we obtain $\sigma _{SH} \sim
10^3$ ($\Omega$ m)$^{-1}$ and $\sigma _{SH}/\sigma_{c} \sim$ 0.4-1
$10^{-4}$, in reasonable agreement with the experimental results.

We have thus presented electrical measurements of the spin-Hall
effect in a diffusive metallic conductor (aluminum), and have
obtained the spin-Hall conductivity. We have studied spin
injection and spin relaxation with both reference spin-transistors
and spin Hall crosses, obtaining consistent results.

This work demonstrates new means to study spin-related phenomena.
It shows that the spin Hall effect can be used to directly measure
spin polarized currents without the need of ferromagnets. Spin
precession does not modify the spin Hall voltage for magnetic
fields perpendicular to the substrate, as in the configuration
discussed here. However, preliminary measurements with a magnetic
field applied parallel to the Al strip suggest the generation of a
spin Hall voltage that is a consequence of out-of-plane spins
induced by spin precession.

\vspace{5mm}

\noindent{\small{*Present address: Francis Bitter Magnet
Laboratory, Massachusetts Institute of Technology, Cambridge, MA
02139, USA}

\vspace{5mm}

\noindent \textbf{Supplementary Information} is linked to the
online version of the paper at www.nature.com/nature.

\vspace{5mm}

\noindent \textbf{Acknowledgments} We thank L. DiCarlo, H.A.
Engel, D.J. Monsma and W.D. Oliver for a critical reading of the
manuscript. This research was supported in part by the US National
Science Foundation and the US Office of Naval Research.

\vspace{5mm}

\noindent \textbf{Author Information} Correspondence and request
for materials should be addressed to S.O.V. (sov@mit.edu).

\end{document}